\documentclass[final,1p,times,nonatbib]{elsarticle}

\pdfoutput=1

\makeatletter
\def\ps@pprintTitle{%
  \let\@oddhead\@empty
  \let\@evenhead\@empty
  \let\@oddfoot\@empty
  \let\@evenfoot\@oddfoot
}
\makeatother

\usepackage[T1]{fontenc}

\usepackage[utf8]{inputenc}

\usepackage[autostyle]{csquotes}

\usepackage{multirow}
\usepackage{graphicx}
\usepackage{booktabs}
\usepackage{subcaption}
\usepackage{geometry}

\usepackage[super]{nth}
\usepackage{supertabular,booktabs}
\usepackage{float}
\usepackage{blindtext}

\usepackage{underscore}

\usepackage{url}
\usepackage{soul}           %
\usepackage{color}
\usepackage{colortbl}

\usepackage{booktabs}       %
\usepackage{enumitem}       %
\usepackage{siunitx}       %
\usepackage{pbox}
\graphicspath{{figures/}}
\usepackage{tabularx}
\usepackage{amsmath,amssymb,amsfonts}%
\usepackage{amsthm}%
\usepackage{mathrsfs}%

\usepackage[title]{appendix}%
\usepackage{xcolor}%
\usepackage{textcomp}%
\usepackage{manyfoot}%
\usepackage{booktabs}%
\usepackage{algorithm}%
\usepackage{algorithmicx}%
\usepackage{algpseudocode}%
\usepackage{listings}%

\usepackage[hidelinks]{hyperref}

\makeatletter
\renewcommand{\sectionautorefname}{\S\@gobble}
\renewcommand{\subsectionautorefname}{\S\@gobble}
\renewcommand{\subsubsectionautorefname}{\S\@gobble}
\makeatother

\usepackage[
backend=biber,
style=apa,
natbib=true,
sorting=nyt]{biblatex}

\usepackage{orcidlink}

\definecolor{LightGray}{gray}{0.9}
\definecolor{grayish}{rgb}{0.95, 0.95, 0.95}

\usepackage{glossaries}
\usepackage{acro}[=v2]

\makeglossaries

\ExplSyntaxOn
\prop_new:N \l__acro_foreign_format_prop
\ExplSyntaxOff

\DeclareAcronym{dlt}{
    short = {DLT},
    long = {Distributed Ledger Technology}
}

\DeclareAcronym{dlts}{
    short = {DLTs},
    long = {Distributed Ledger Technologies}
}

\DeclareAcronym{pow}{
    short = {PoW},
    long = {Proof of Work}
}

\DeclareAcronym{pos}{
    short = {PoS},
    long = {Proof of Stake}
}

\DeclareAcronym{evm}{
    short = {EVM},
    long = {Ethereum Virtual Machine}
}

\DeclareAcronym{esg}{
    short = ESG,
    long = {Environmental, Social, and Governance}
}

\DeclareAcronym{ai}{
    short = {AI},
    long = {Artificial Intelligence}
}

\DeclareAcronym{nlp}{
    short = {NLP},
    long = {Natural Language Processing}
}

\DeclareAcronym{ner}{
    short = {NER},
    long = {Named Entity Recognition}
}

\DeclareAcronym{qa}{
    short = {QA},
    long = {Question Answering}
}

\DeclareAcronym{bert}{
    short = BERT,
    long = Bidirectional Encoder Representations from Transformers
}

\DeclareAcronym{llms}{
    short = {LLMs},
    long = {Large Language Models}
}

\DeclareAcronym{llm}{
    short = {LLM},
    long = {Large Language Model}
}

\DeclareAcronym{gpt}{
    short = {GPT},
    long = {Generative Pre-trained Transformer}
}

\DeclareAcronym{tradfi}{
    short = TradFi,
    long = Traditional Finance
}

\DeclareAcronym{defi}{
    short = DeFi,
    long = Decentralized Finance
}

\DeclareAcronym{amm}{
    short = AMM,
    long = Automated Market Maker
}

\DeclareAcronym{dex}{
    short = DEX,
    long = Decentralized Exchange
}

\DeclareAcronym{dexs}{
    short = DEXs,
    long = Decentralized Exchanges
}

\newglossaryentry{nlp}{
    name={\acl{nlp} (\acs{nlp})},
    description={Represents a branch of \ac{ai}\footnote{https://cso.kmi.open.ac.uk/topics/artificial\_intelligence} and linguistics\footnote{https://cso.kmi.open.ac.uk/topics/linguistics} that focuses on understanding, interpreting, and processing human language (in spoken and written form) in a way that is both meaningful and useful. While \ac{nlp} encompasses both spoken and written language processing, this paper specifically focuses on written text analysis through the application of language models for an \ac{ner} task}
}

\newglossaryentry{blockchain}{
    name={Blockchain},
    description={
        A popular form of \ac{dlt} where data entries, called 
        \emph{blocks}, are linked and secured using cryptography. 
        \emph{Bitcoin} is one notable blockchain application \citep{Nakamoto2008Bitcoin:System}}
}

\newglossaryentry{consensus}{
    name={Consensus Mechanism},
    description={Rules and procedures that determine how network participants agree on the validity of transactions and maintain the blockchain. Examples include \ac{pow} used by Bitcoin \citep{Nakamoto2008Bitcoin:System}, \ac{pos} used by Ethereum \citep{Buterin2014Ethereum:Platform.} and Hedera \citep{Baird2018Hedera:Council}, etc}
}

\newglossaryentry{hashgraph}{
    name={Hashgraph},
    description={
        An alternative to blockchain that implements a Directed Acyclic Graph (DAG) data structure, a graph of hashes (\blockquote{hashgraph}), and its own type of consensus algorithm \citep{Baird2018Hedera:Council} while retaining the core principles of decentralized validation and record-keeping}
}

\newglossaryentry{languagemodel}{
    name={Language Model},
    description={Statistical models trained on large text corpora. Modern architectures of language models, like \ac{bert} and \ac{gpt}, first undergo pre-training, which is a process where they learn general language patterns and representations from vast amounts of unlabeled text}
}

\newglossaryentry{bert}{
    name={\ac{bert}},
    description={\acl{bert} (\ac{bert}) \citep{Devlin2019BERT:Understanding}. Pre-trained language model that can be fine-tuned to perform downstream tasks with labeled datasets (supervised learning)}
}

\newglossaryentry{gpt}{
    name={\ac{gpt}},
    description={\acl{gpt} (\acs{gpt}) \citep{Radford2018ImprovingPre-Training,Radford2019LanguageLearners}. Pre-trained language model can perform tasks through prompting mechanisms \citep{Lu2022FantasticallySensitivity} like zero-shot and few-shot}
}

\newglossaryentry{sequencetagging}{
    name={Sequence Tagging},
    description={The process of assigning categorical labels to individual elements (tokens) in a text sequence, where each token's label depends on both its content and context. Common sequence tagging tasks include \ac{ner}, Part of Speech tagging (labeling words as nouns, verbs, etc.), Chunking (identifying meaningful phrase groups), Semantic role labeling (determining word functions in sentences), etc}
}

\newglossaryentry{ner}{
    name={\acl{ner} (\acs{ner})},
    description={A sequence tagging task that involves identifying and categorizing specific information within text into predefined groups or categories}
}

\newglossaryentry{fine-tuning}{
    name={Fine-tuning},
    description={Supervised learning approach for task-specific adaptation for \ac{bert}-based models, where a pre-trained \ac{bert}-based model undergoes additional training on a task-specific labeled dataset. For example, a fine-tuned language model for an \ac{ner} task would analyze the sentence \blockquote{Bitcoin uses Proof-of-Work consensus mechanism} and identify \blockquote{Bitcoin} as a \texttt{Blockchain Name} and \blockquote{Proof-of-Work} as a \texttt{Consensus} mechanism. This identification process relies on the model's ability to understand context and recognize patterns in text sequences based on the labeled data or prompt provided}
}

\newglossaryentry{zeroshot}{
    name={Zero-shot Learning},
    description={Prompting mechanisms where \ac{gpt}-based language models attempts tasks without any examples}
}

\newglossaryentry{fewshot}{
    name={Few-shot Learning},
    description={Prompting mechanisms where \ac{gpt}-based language models receive a small number of examples within the input prompt for it to perform a task}
}

\begin{document}

\begin{frontmatter}

\title{Evolution of ESG-focused DLT Research: An NLP Analysis of the Literature}

\author[inst1,inst2,inst3]{Walter Hernandez Cruz\orcidlink{0000-0002-0253-8117}\corref{cor1}}
\ead{walter.hernandez.18@ucl.ac.uk}

\author[inst1,inst2,inst3]{Kamil Tylinski\orcidlink{0009-0005-4611-2336}}
\ead{kamil.tylinski.16@ucl.ac.uk}

\author[inst1]{Alastair Moore\orcidlink{0000-0002-3726-6473}}
\ead{a.p.moore@ucl.ac.uk}

\author[inst1]{Niall Roche\orcidlink{0000-0001-6189-726X}}
\ead{n.roche@ucl.ac.uk}

\author[inst1,inst2,inst3]{Nikhil Vadgama\orcidlink{0000-0002-3303-646X}}
\ead{nikhil.vadgama@ucl.ac.uk}

\author[inst1,inst4]{Horst Treiblmaier\orcidlink{0000-0002-0755-5223}}
\ead{horst.treiblmaier@modul.ac.at}

\author[inst1,inst5]{Jiangbo Shangguan\orcidlink{0009-0000-6347-0403}}
\ead{jiangbo.shangguan@pku.org.uk}

\author[inst2,inst3]{Paolo Tasca\orcidlink{0000-0002-5460-5940}}
\ead{p.tasca@exp.science}

\author[inst1,inst2,inst3]{Jiahua Xu\orcidlink{0000-0002-3993-5263}}
\ead{jiahua.xu@ucl.ac.uk}

\cortext[cor1]{Corresponding author}

\affiliation[inst1]{organization={Centre for Blockchain Technologies, University College London},%
            }

\affiliation[inst2]{organization={UK Centre for Blockchain Technologies},%
            }

\affiliation[inst3]{organization={Exponential Science},%
            }

\affiliation[inst4]{organization={Modul University Vienna},%
            }

\affiliation[inst5]{organization={HSBC Business School, Peking University},%
            }

\begin{abstract}

\ac{dlt} faces increasing environmental scrutiny, particularly concerning the energy consumption of the \ac{pow} consensus mechanism and broader \ac{esg} issues. However, existing systematic literature reviews of \ac{dlt} rely on limited analyses of citations, abstracts, and keywords, failing to fully capture the field's complexity and \ac{esg} concerns. We address these challenges by analyzing the full text of 24,539 publications using \ac{nlp} with our manually labeled \ac{ner} dataset of 39,427 entities for \ac{dlt}. This methodology identified 505 key publications at the \ac{dlt}/\ac{esg} intersection, enabling comprehensive domain analysis. 

Our combined \ac{nlp} and temporal graph analysis reveals critical trends in \ac{dlt} evolution and \ac{esg} impacts, including cryptography and peer-to-peer networks research's foundational influence, Bitcoin's persistent impact on research and environmental concerns (a \blockquote{Lindy effect}), Ethereum's catalytic role on \ac{pos} and smart contract adoption, and the industry's progressive shift toward energy-efficient consensus mechanisms.

Our contributions include the first \ac{dlt}-specific \ac{ner} dataset addressing the scarcity of high-quality labeled \ac{nlp} data in blockchain research, a methodology integrating \ac{nlp} and temporal graph analysis for large-scale interdisciplinary literature reviews, and the first \ac{nlp}-driven literature review focusing on \ac{dlt}'s \ac{esg} aspects.

\end{abstract}


\end{frontmatter}

\section{Introduction}
Emerging technologies are facing increased scrutiny regarding their energy consumption and broader ecological impacts, including the use of vital resources such as water, precious metals, and synthetic compounds \citep{Platt2021a, Simone2022EconomicInvestigation,Sun2024ESGReturns}. This growing environmental awareness necessitates evaluating technological advancements through their ecological footprint, with \ac{dlt} being no exception. While \ac{dlt} offers promising features like record immutability and decentralization, it also presents challenges, particularly the high energy consumption of specific consensus algorithms. For instance, Bitcoin's \ac{pow} \citep{Nakamoto2008Bitcoin:System}, designed to prevent Sybil attacks, in which malicious actors pretend to be multiple users, has significant energy requirements \citep{Ibanez2023BitcoinsExpansion}. 

The rapid evolution and interest in \ac{dlt} have led to a growing number of publications and intensified scrutiny of its ecological footprint. Yet, persistent misconceptions about its energy consumption patterns have complicated objective assessment. 
Additionally, the complexity of \ac{dlt} extends well beyond environmental considerations. \ac{dlt} spans breakthroughs in security, cryptography, network design, and diverse applications, leading to a rapidly expanding body of research. Therefore, traditional systematic literature review methodologies and manual literature analysis techniques that primarily rely on citation metrics, abstract analyses, and database term searches demonstrate increasing insufficiency in capturing the multidimensional nature and full scope of the rapidly evolving \ac{dlt} domain.

We address this complexity by turning to \ac{nlp}, a subfield of \ac{ai} and linguistics. \ac{nlp} offers powerful tools to examine the growing body of \ac{dlt} literature, from academic articles to industry whitepapers. Among its various applications, we focus on \ac{ner} to identify specific \ac{dlt} technologies (e.g., \ac{pow} and \ac{pos}) and their \ac{esg} implications (e.g., \blockquote{energy consumption} and \blockquote{computational power}) within our dataset. This approach allows us to understand the shifts in research across different \ac{dlt} technologies and their environmental impacts.

Building on precedents from the biomedical field, where ontologies or hierarchical taxonomies are used to create \ac{ner} datasets for \ac{nlp}-driven literature mining and review \citep{Spasic2005TextText, Huang2020BiomedicalDevelopment, Nabi2022ContextualOntologies, Chang2016DevelopingPipeline, Mcentire2016ApplicationDevelopment, Alsheikh2022TheDiseases}, we have successfully transferred these methods to the \ac{dlt} domain. Unlike previous systematic literature reviews that rely on citation measures, abstract analyses, keywords, or database term searches, our approach examines the entire body of each publication. This allows us to capture all relevant entities under top-level categories representing the whole \ac{dlt} field, removing the need for exhaustive database term searches to generate a dataset representing the \ac{dlt} field and enabling us to work at scale. By mapping each publication’s tokens to the hierarchical \ac{dlt} taxonomy from \citet{Tasca2019AClassification}, which we extended to include \ac{esg} considerations and other categories, we detected thematic shifts in both academic research and industry publications (e.g., whitepapers). As a result, we compiled a more inclusive dataset that reflects the \ac{dlt} field’s overall evolution while preserving the methodological rigor demonstrated in biomedical literature studies.

Our research makes the following contributions:

\begin{itemize}

    \item Creating a manually labeled \ac{ner} dataset of 39,427 named entities for twelve top-level \ac{dlt}'s taxonomy categories. To the best of our knowledge, this is the first \ac{ner} dataset explicitly designed for \ac{dlt}. Our dataset aims to address the \ac{nlp} data scarcity issue in blockchain research. By making it openly available, we help overcome common limitations like high labeling costs and complex access to high-quality labeled \ac{nlp} data in the \ac{dlt} field.\footnote{The dataset is available at: https://huggingface.co/datasets/ExponentialScience/ESG-DLT-NER}
    
    \item Presenting a methodological framework for executing an \ac{nlp}-driven systematic literature review at the intersection of domains, in this case, \ac{dlt} and \ac{esg} research. Our methodology combines approaches from different publications and fields for domain-specific literature text mining and systematic literature reviews.

    \item Conducting the first \ac{nlp}-driven systematic literature review for the \ac{dlt} field that places a special emphasis on \ac{esg} aspects, deriving empirical insights into the evolution of the literature.

\end{itemize}

Our work, particularly our \ac{ner} dataset, can potentially support future \ac{nlp} data-driven studies in the \ac{dlt} field. Additionally, it represents a foundation for future research to improve automated systematic literature review processes at scale, capable of capturing the intrinsic dependencies and evolution of concepts related to the intersection of fields.

\section{Related work}

\paragraph{Systematic Reviews in \ac{dlt}}

Previous literature reviews have extensively explored blockchain applications in various sectors \citep{Casino2019AIssues, Zheng2018BlockchainSurvey}. These reviews, however, differ in scope and depth compared to our systematic review, particularly in terms of the number of analyzed articles. Previous studies have mainly focused on blockchain's role in decentralization and privacy and analyzed trends of centralization in decentralized systems such as Bitcoin and Ethereum \citep{Sai2021TaxonomyReview}. \citet{Spychiger2021UnveilingBlockchain} deconstructed 107 blockchain technologies using a specific taxonomy, emphasizing consensus mechanisms and cryptographic primitives. Our work, in contrast, provides a broader perspective on the evolution of \ac{dlt}, including its \ac{esg} implications, and uses a bigger corpus.

\paragraph{\ac{esg} Implications}

\citet{Bilal2014ANetworks, Mengelkamp2018AMarkets, Poberezhna2018AddressingFinance, Schulz2021LeveragingFund, Wu2022ConsortiumCrowdsensing, Jiang2022Blockchain-basedReporting,Richardson2020e} have explored blockchain's potential in energy management, environmental sustainability, carbon trading, and transparent reporting. Our study extends these approaches by examining the intersection of \ac{esg} and \ac{dlt} through a literature analysis.

\paragraph{\ac{nlp} Applications}

Studies have demonstrated \ac{nlp}'s effectiveness in automated \ac{esg} scoring \citep{AlikSokolov2021BuildingScoring} and opinion summarization \citep{Dubey2023SmartBlockchain}. Outside \ac{dlt}, systematic reviews using \ac{nlp}-driven methodologies, particularly in medical genomics \citep{Alsheikh2022TheDiseases}, have employed database term searches and abstract-based filtering, contrasting with our approach that uses a taxonomy, labeled \ac{ner} dataset, and language model to compile and filter a more inclusive corpus representing the \ac{dlt} field and from which we derive insights.

\subsection{Literature reviews of \ac{dlt} application in industry sectors}

\paragraph{Supply Chain Applications}
Blockchain applications in supply chain management have been extensively studied \citep{Wannenwetsch2023BlockchainChallenges,Casella2023CasesReview,Ibanez2025Triple-EntryKin}, focusing on traceability, transparency, and process automation. These reviews highlight blockchain's role in enhancing supply chain visibility and reducing intermediary dependencies.

\paragraph{Internet of Things (IoT) Integration}
Research examining blockchain-IoT integration \citep{Dai2019BlockchainSurvey,Lo2019AnalysisReview,Alladi2019BlockchainReview,Sallam2021BlockchainReview} has emphasized security enhancement, data integrity, and decentralized device management. Studies have particularly focused on scalability challenges \citep{Liu2025DynaShard:Management} and consensus mechanisms suitable for resource-constrained IoT devices.

\paragraph{Healthcare Systems}
Literature reviews in healthcare applications \citep{Ghosh2020AConsumption,Khatoon2020AManagement,Hussien2019ADirection} have explored blockchain's potential for secure health records management, clinical trial data integrity, and pharmaceutical supply chain tracking.

\paragraph{Financial Sector}
Blockchain-based shared ledgers have long been studied for their potential to impact accounting, auditing, and finance. In particular, the evolution of \ac{defi} has generated systematic reviews \citep{Shah2023AProtocols,Gramlich2023AAvenues,Chen2020BlockchainModels,Cousaert2021} examining blockchain's potential to reduce financial intermediation. This potential is exemplified through lending protocols \citep{Xu2022d}, stablecoins \citep{Ante2023AStablecoins}, and \acp{dex} \citep{xu2021dexAmm}, which operate via smart contracts \citep{Adams2021UniswapCore} or at the protocol level of a blockchain \citep{HernandezCruz2025AMM-basedLedger,Peduzzi2021}. In particular, stablecoins \citep{Ante2023AStablecoins} serve as the critical infrastructure bridging \ac{tradfi} and \ac{defi}, though this interconnection can propagate market shocks between both \citep{HernandezCruz2024NoBank}.

\section{Methodology} \label{sec:methodology}

\begin{figure*}[hbt]

     \centering

     \includegraphics[width=\textwidth]{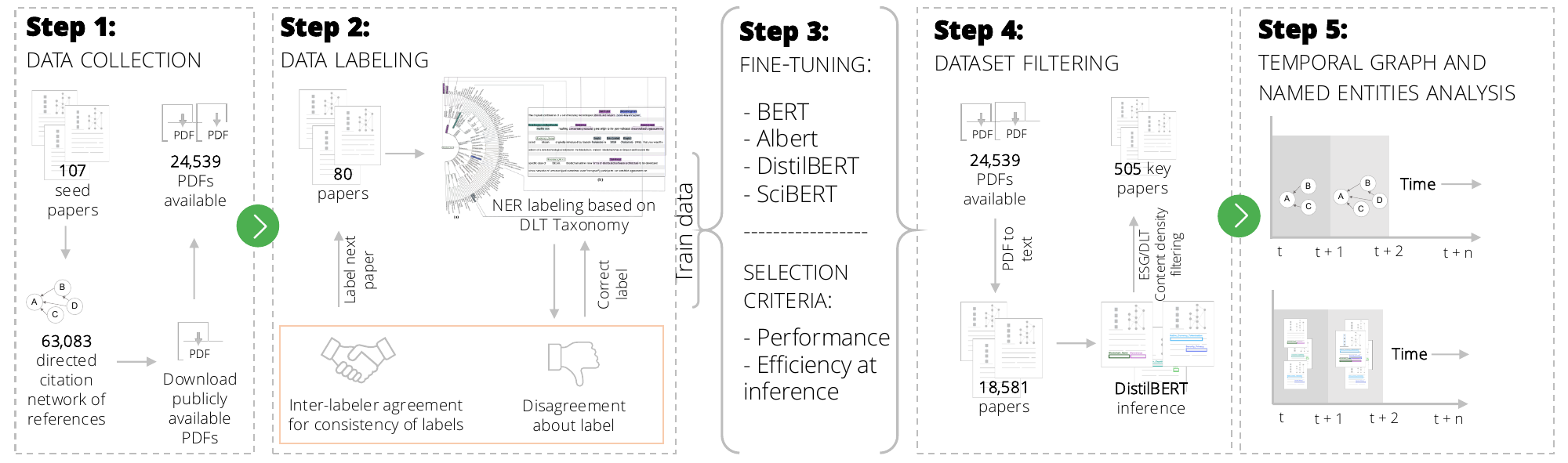}
     
     \caption{Methodology for the systematic literature review of ESG/DLT publications evolution using \ac{ner} for content filtering, for which temporal graph and named entities (representing specific \ac{dlt} technologies, e.g., \ac{pow} and \ac{pos}) analysis is carried.}
     \label{fig:methodology}
 \end{figure*}

\begin{figure*}[ht!]
     \centering
      \includegraphics[width=\textwidth]{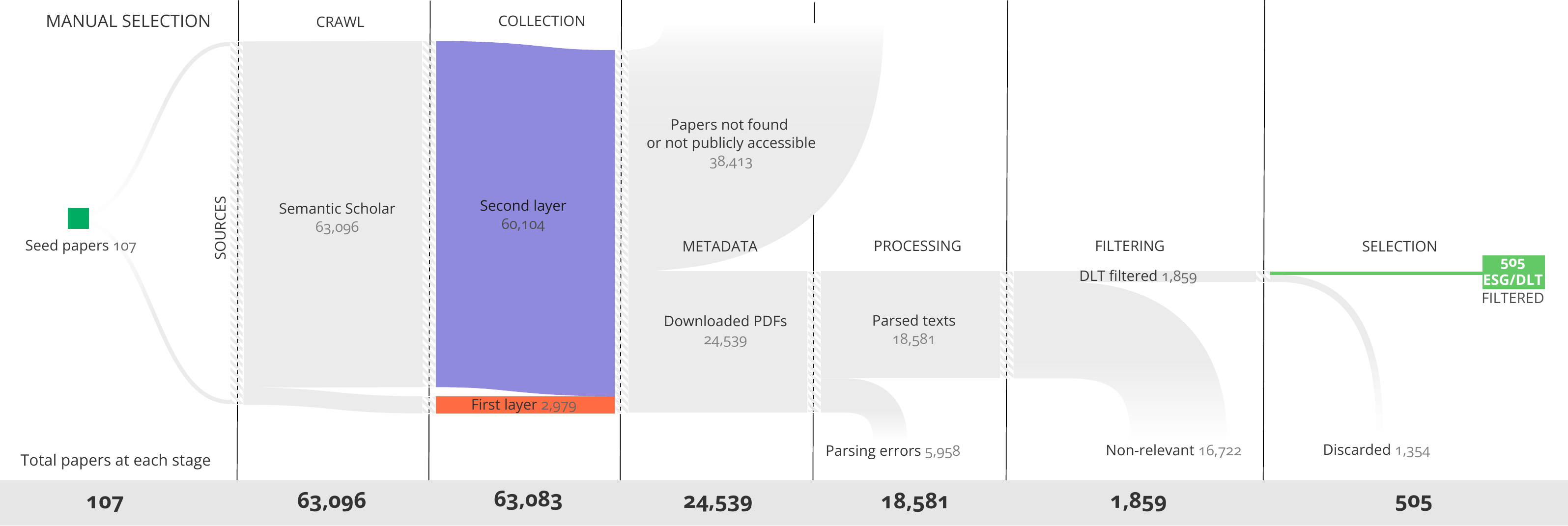}

      \caption{Processing pipeline for collecting and filtering papers in the review. The total number of papers present at each stage of processing is shown. See \protect{\autoref{tab:labelsDLT}} for the description of the labels in the corpus.}

     \label{fig:workflow}
 \end{figure*}

\paragraph{Dataset construction and analysis approach}
Ontologies, specifically hierarchical taxonomies, are pivotal in developing \ac{ner} datasets for text mining \citep{Spasic2005TextText, Huang2020BiomedicalDevelopment, Nabi2022ContextualOntologies, Chang2016DevelopingPipeline, Mcentire2016ApplicationDevelopment, Alsheikh2022TheDiseases}. For example, the GENIA corpus \citep{Kim2003GENIABio-textmining}, a \ac{ner} dataset from 2,000 biological abstracts, employs the GENIA ontology's hierarchical tree structure of 47 biological entities, including top-level categories such as biological source and substance to facilitate text mining in biomedical literature. Similarly, the Human Phenotype Ontology is used to create and expand \ac{ner} datasets in biomedicine \citep{Lobo2017IdentifyingRules, Huang2020BiomedicalDevelopment}. \citet{Alsheikh2022TheDiseases, Chang2016DevelopingPipeline, Mcentire2016ApplicationDevelopment} further demonstrate the use of ontology-based \ac{ner} datasets for domain-specific literature text mining.

Learning from these biomedical field precedents, our methodology (\autoref{fig:methodology}) for \ac{nlp}-based text mining and filtering (\autoref{fig:workflow}) in the \ac{dlt} field employs a hierarchical taxonomy \citep{Tasca2019AClassification} to manually annotate a \ac{ner} dataset of 80 full-text papers: 46 systematically reviewed publications of \ac{dlt}'s sustainability \citep{Eigelshoven2020PublicAlgorithms} and 34 manually selected for their relevance to the ESG/DLT theme. 

\paragraph{Two-tier analysis strategy}
Our analysis followed a two-tier approach to maximize the use of available data:

1. \textit{Broad analysis using metadata}: Keyword extraction \citep{Rose2021ARAKE} and topic modeling \citep{Asmussen2019SmartReview} provide a broad view of the main themes for a literature review. We analyzed the complete corpus of over 60,000 papers using their metadata (e.g., titles and keywords), regardless of full-text availability. This approach allows us to construct a comprehensive keywords graph that captures the broad evolution of the \ac{dlt} field. While many papers (38,413) were behind paywalls or otherwise not easily accessible, their metadata still contributes to our understanding of the field's development.

2. \textit{Deep analysis using full text}.
We conducted a more detailed analysis of the subset of 24,539 papers where full text was available through open-access journals, conferences, or public repositories. We narrowed the number of publications for our analysis by fine-tuning a transformer-based language model for a \ac{ner} task for ESG/DLT content density corpus filtering. This method allows us to categorize technologies in \ac{dlt}, such as different \texttt{Consensus} mechanisms (e.g., \acl{pow}, \acl{pos}) and their \texttt{\ac{esg}} implications, like \enquote{energy consumption} and \enquote{computational power}, that would not usually emerge as single topics or keywords for a literature review. This deeper analysis reveals nuanced connections that may not arise from metadata alone.

\paragraph{Temporal graph and named entities analysis}
The final stage involves a temporal graph and named entities analysis to track the evolution of \ac{dlt} technologies and their \ac{esg} implications over time. This combines insights from the broad metadata analysis and the detailed full-text examination, providing a comprehensive view of the field's development.

\subsection{Data collection} \label{subsec:DataCollection}

We constructed a directed citation network using a two-level expansion method: beginning with seed papers, collecting their references, and then gathering the references of those references. This two-level restriction helps maintain a thematic focus on \ac{esg}/\ac{dlt}. The resulting network comprises 63,083 publications. Through Semantic Scholar's metadata for each publication (see \autoref{fig:methodology}), we retrieved 24,539 full-text PDFs that are openly accessible through open-access journals, institutional repositories, preprint servers (e.g., arxiv), or publisher websites (see \autoref{fig:methodology} and \autoref{fig:workflow}).

The key benefit of using seed papers to build a citation network for a systematic literature review is the ease of expanding and updating the literature review by selecting seminal publications from the academic literature. The seed papers for our directed citation network were selected from two sources:
\begin{enumerate}
    \item 89 papers from \citep{Eigelshoven2020PublicAlgorithms}, reviewing sustainability in popular \ac{dlt} consensus algorithms.
    \item 17 publications (2018-2022) with at least three citations each, chosen to update the corpus with more current research relevant to the ESG/DLT intersection \citep{Platt2021a, Kohli2022AnSolutions, Ante2021BitcoinsLayer, Sedlmeir2020TheMyth, Fernando2021BlockchainCompliance, Masood2018ConsensusEnvironment, Ghosh2020AConsumption, Eshani2021AnIt, Cole2018ModelingAlgorithms, Lucey2021AnICEA, Sapkota2020BlockchainPrices, Bada2021TowardsConsumption, Denisova2019BLOCKCHAINCONSUMPTION, Schinckus2020CRYPTO-CURRENCIESCONSUMPTION, Sedlmeir2021RecentConsumption, Powell2021AwarenessBlockchain, Alofi2022OptimizingFormulation}.     
\end{enumerate}

\subsection{\acs{ner} task for literature mining} \label{subsec:ner-literature-mining}

We considered using \acp{llm} for our \ac{ner} task. An unsupervised approach like zero-shot \ac{ner} with \acp{llm} would have been more cost-effective in not requiring manually labeled data. However, at the time of this writing, domain-specific \ac{ner} tasks often perform better with supervised learning models than with current \acp{llm} \citep{Li2023AreTasks,Hu2024ImprovingEngineering}. This performance difference arises from fundamental architectural distinctions: \ac{bert}-based models generally outperform decoder-only \acp{llm}, such as OpenAI's ChatGPT and Google Deepmind's Gemini, on token-level tasks like \ac{ner}, due to their bidirectional encoder architecture \citep{Devlin2019BERT:Understanding}. Specifically, \ac{bert} processes text by simultaneously considering both left and right context \citep{Devlin2019BERT:Understanding}, enabling it to capture fine-grained contextual information around each token. This is a crucial capability for \ac{ner}, where entity recognition often depends on the complete contextual surrounding of a word. 

In contrast, decoder-only architectures, like \ac{gpt}, process text unidirectionally (left to right) \citep{Radford2018ImprovingPre-Training,Radford2019LanguageLearners}, limiting their ability to capture the complete contextual information necessary for accurate entity recognition. While decoder-only architectures excel at open-ended question answering, they struggle to match \ac{bert}-based models' \ac{ner} performance \citep{Li2023AreTasks,Hu2024ImprovingEngineering}, which currently represents the state of the art for \ac{ner} tasks\footnote{https://paperswithcode.com/task/named-entity-recognition-ner}. Therefore, we adopted a supervised learning approach, fine-tuning and benchmarking several \ac{bert}-base models, selecting the final model based on its performance and efficiency at inference. This approach provides more granularity for literature mining an entire publication body without the hallucination issues of \acp{llm} (i.e., creation of irrelevant or inconsistent content). However, the trade-off is the requirement for the costly manual labeling of the \ac{ner} dataset.

\subsubsection{Labeling}\label{subsubsec:labeling}

\begin{figure*}[!htb]
   \centering
     \centering
      \includegraphics[width=1\textwidth]{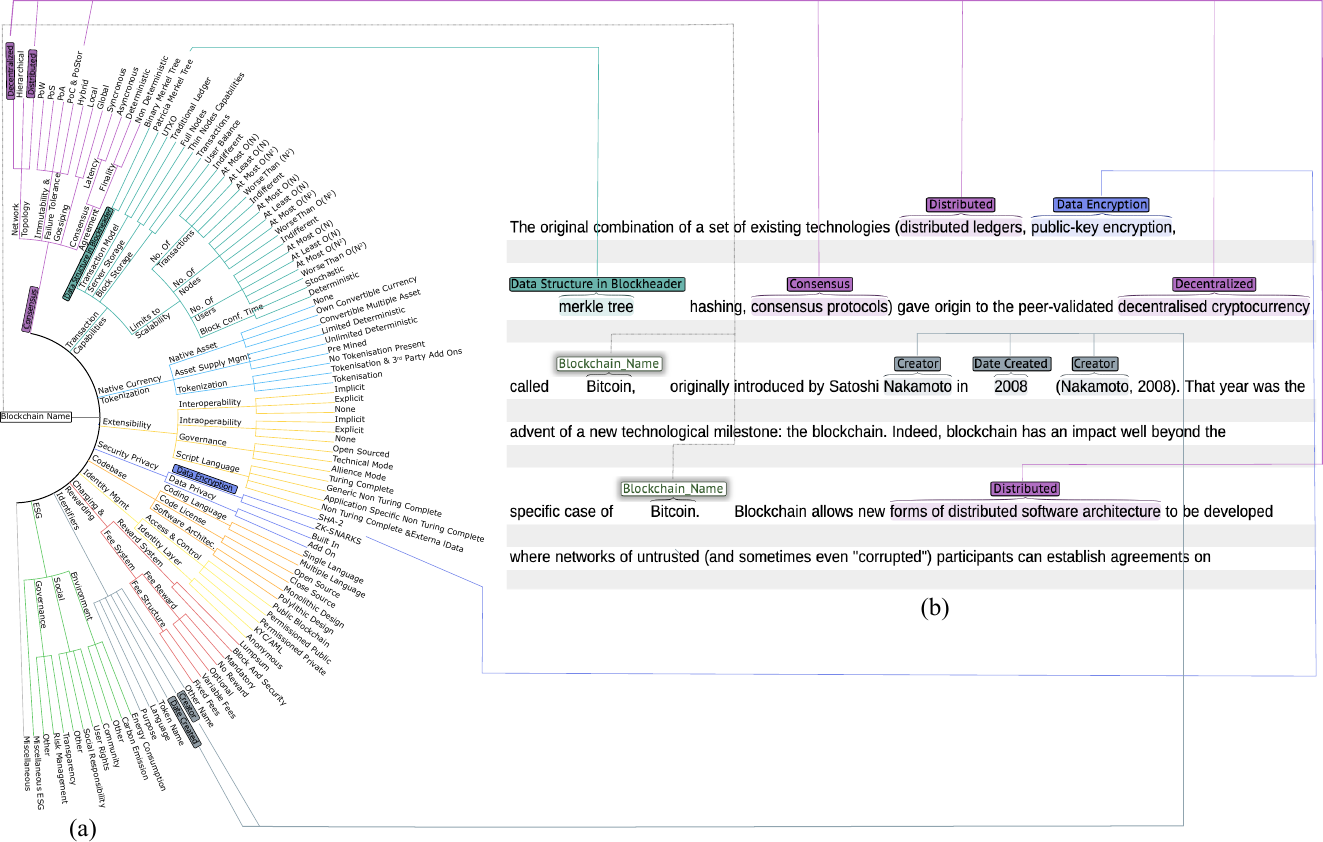}
     \caption{(a) The taxonomy of \citet{Tasca2019AClassification} extended with \texttt{Blockchain Name}, \texttt{ESG}, and \texttt{Miscellaneous} (see \autoref{subsubsec:labeling}) for the purpose of this research. (b) Example of parsed text with the taxonomy label associated with a span of text labeled. The labels used in the paragraph are highlighted in the taxonomy tree.} 
     \label{fig:taxonomy}
 \end{figure*}

\begin{table}[!htbp]
\centering
\footnotesize %
\caption{List of groups of entity types based on the extended taxonomy from ~\citet{Tasca2019AClassification} (see \autoref{fig:taxonomy}).}
\begin{tabular}{p{0.27\textwidth} p{0.70\textwidth}} %
\toprule
\textbf{Group entities} & 
\textbf{Description} \\
\midrule
\texttt{Blockchain Name} & The name of a blockchain system (e.g., Bitcoin, Ethereum, XRP Ledger), but also including other types of \acp{dlt}, such as Hedera, IOTA, and Nano. \\
\texttt{Consensus} & Rules and mechanisms to ensure the immutability of transaction records (e.g., \ac{pow}, \ac{pos}, Blockchain, Hashgraph, Gossiping, Decentralized, Centralized, Distributed, Hierarchical, Latency, and Finality). \\
\texttt{Transaction Capabilities} & Information related to the details of transactions, such as Data Structure in the Blockheader, Transaction Model, Server Storage, Block Storage, and Limits to Scalability. \\
\texttt{Native Currency Tokenization} & Asset classes to pay for transactions within a blockchain system (e.g., BTC, ETH, XRP, HBAR). \\
\texttt{Extensibility} & Capabilities of Interoperability, Intraoperability, Governance (e.g., Decentralized Autonomous Organization (DAO), Alliance Mode, Open-source community mode), and Script Language of a blockchain ecosystem (e.g., Turing complete, non-Turing complete). \\
\texttt{Security Privacy} & Cryptographic methods to ensure data privacy, encryption, and security in a blockchain ecosystem (e.g., private key, public key, hashing, Sybil attack, 51\% attack). \\
\texttt{Codebase} & Coding Language (e.g., Solidity, Vyper, Rust, JavaScript), Code License (e.g., GNU General Public License (GPL), MIT License, Apache License), and Software Architecture of a blockchain ecosystem (e.g., Monolithic, Polylithic). \\
\texttt{Identity Management} & Attributes to identify participants and their system access level. \\
\texttt{Charging \& Rewarding System} & Cost and reward models for the operation and maintenance of blockchain systems (e.g., transaction fees, mining rewards, fee rewards, fee system, fee structure). \\
\texttt{Identifiers} & Information related to the token names (e.g., USDC, USDT), creators (e.g., Satoshi Nakamoto, Ripple, Leemon Baird, Circle, Coinbase), and purpose (e.g., stablecoins, utility token, security token). \\
\texttt{ESG} & Entities relevant to \ac{esg} issues regarding Blockchain (e.g., energy consumption, computing power). \\
\texttt{Miscellaneous} & Miscellaneous entities that are ambiguous in a given context and are relevant for the \ac{dlt} topic but are not captured by any of the above categories. \\
\bottomrule
\end{tabular}
\label{tab:labelsDLT}
\end{table}

We manually annotated 80 publicly available publications using the brat tool \citep{Stenetorp2012BRAT:Annotation} and argilla\footnote{https://github.com/argilla-io/argilla}, following the extended taxonomy framework of \citet{Tasca2019AClassification} (\autoref{fig:taxonomy}). This taxonomy provides a hierarchical structure of \ac{dlt} technology components, with each principal component (e.g., \texttt{Consensus}) divided into sub-components (e.g., \texttt{Gossiping}) and further into sub-sub-components if needed (e.g., \texttt{Local}). We introduced categories like \texttt{Blockchain Name} to identify specific blockchains. We expanded the initial definition of \texttt{Security Privacy} to label security threats (e.g., \texttt{Sybil attack} and \texttt{51\% attack}). Also, we added the \texttt{Miscellaneous} category for ambiguous contexts (see \autoref{fig:taxonomy} and \autoref{tab:labelsDLT}), following the example of the CoNLL-2003 dataset for a similar category \citep{Tjong2003IntroductionRecognition}. We further extended \citet{Tasca2019AClassification}'s taxonomy to identify sustainability-related concepts referred to in the \texttt{\ac{esg}} criterion (see \autoref{fig:taxonomy}).

\subsubsection{Text analysis/language processing}\label{subsubsec:LabelsProcessing}

We pruned the label hierarchy within the taxonomy for class balance, where specific labels such as \texttt{PoW} are replaced by broader categories such as \texttt{Consensus} to maintain focus on primary taxonomy components (\autoref{fig:taxonomy}). To improve the \ac{ner} model performance, which is sensitive to label consistency \citep{Zeng2021ValidatingAnnotation, Jeong2023ConsistencyRecognition}, we employed a systematic process for enhancing inter-labeler consistency. This process involved correcting inconsistent labeling of entities. For example, \blockquote{Sybil attack} was sometimes labeled as \texttt{Consensus} and, at other times, as \texttt{Security Privacy}. We resolved this inconsistency following each labeler's approval and using programmatic cleaning to ensure consistency for non-context-dependent labels. For \blockquote{Sybil attack} specifically, we decided to label it as \texttt{Security Privacy}.

We applied text resampling for overlapping named entities that could fit into multiple categories, such as belonging to \texttt{Blockchain Name} and \texttt{Native Currency Tokenization}. This process involves duplicating text and assigning distinct entities to each copy, enhancing the capture of rare entities. This resampling strategy is beneficial, especially for datasets of modest size \citep{Wang2022Sentence-LevelRecognition}, improving model performance by accommodating diverse entity categories. Additionally, the duplication of training data enhances a language model's ability to learn from limited examples \citep{Muennighoff2023ScalingModels,Charton2024EmergentExamples}.

\subsubsection{Dataset filtering} \label{subsubsec:DatasetFiltering}

We applied a percentile-based filtering process to the corpus of publications based on the \ac{esg} and \ac{dlt} classified named entities within the corpus. This method selects publications with substantial \ac{dlt} and \ac{esg} content, using a threshold percentile to exclude marginally relevant papers. Seed papers were included to maintain foundational references. The filtering is represented as:

\begin{equation}
F = \{P_i : N(P_i) \geq \text{perc}_{10}(N(P))\} \;\cap\; \{P_i : D_{DLT}(P_i) \geq \text{perc}_{90}\} \;\cap\; \{P_i : D_{ESG}(P_i) \geq \text{perc}_{70}\} \;\cup\; S
\end{equation}

where $F$ is the final set of papers, $P_i$ is an individual paper, $N(P_i)$ represents the total token count for paper $i$ (with a minimum threshold at the \nth{10} percentile to exclude potentially corrupted or incomplete documents), $D_{DLT}(P_i)$ and $D_{ESG}(P_i)$ are the \ac{dlt} and \ac{esg} content densities respectively, and $S$ is the set of seed papers. We define the percentile cut-offs for the \ac{dlt} and \ac{esg} content densities by manually reviewing the ratio of \ac{ner} labeled tokens to all tokens in the dataset.

Our processing, filtering, and selection methodology involved (see \autoref{fig:workflow}):

\begin{enumerate}
    \item Processing: Excluding papers below the \nth{10} percentile ($N(P_i) \geq \text{perc}_{10}(N(P))$) in the total token count as a quality control mechanism, eliminating documents that may have been improperly processed during PDF-to-text conversion or are too brief to contain meaningful analysis (e.g., below 100 tokens). We named these discarded papers as \blockquote{parsing errors} in \autoref{fig:workflow}. 
        
    \item \ac{dlt} content filter ($D_{DLT}(P_i) \geq \text{perc}_{90}$): Computing \ac{dlt} content density and retaining papers above the \nth{90} percentile, ensuring a strong focus on \ac{dlt} topics. 
    
    \item Filtering for at least the \nth{70} percentile in \ac{esg} content density to confirm relevance to \ac{esg} ($D_{ESG}(P_i) \geq \text{perc}_{70}$).
    
\end{enumerate}

Finally, we manually reviewed the selected filtered publications to validate the accuracy of their \acs{esg}/\acs{dlt} content density and relevance.

\subsection{Network graphs and entities evolution} \label{subsec:NetworkGraphsAndEntitiesEvolution}

\paragraph{Keywords network}
We constructed a graph $G_{keywords}(V_{keywords}, E_{keywords})$ based on the full corpus of over 60,000 metadata records, with vertices $V_{keywords}$ as individual keywords and edges $E_{keywords}$ representing keywords co-occurrence within the same paper. We analyzed degree centrality with one-year rolling windows to observe the evolution and significance of specific keywords over time, providing insights into how research from these keywords representing topics was foundational for the \ac{dlt} field and facilitated the emergence of innovations.

\paragraph{\ac{esg}/\ac{dlt} network}
We analyzed the \ac{esg}/\ac{dlt} content density filtered citation network as $G(V, E)$, with papers as vertices $V$ and citations as edges $E$. We performed temporal graph analysis using one-year time windows $W_1, W_2, \ldots, W_n$, following the rolling window approach from \citet{Hoadley2021ANetwork, Steer2020Raphtory:Graphs, Steer2024Raphtory:Python}. For each window $W_i$, we created a subgraph $G_i(V_i, E_i)$ and analyzed nodes, edges, average degree, and authority scores (using the HITS algorithm \citep{Kleinberg2011AuthoritativeEnvironment}) to determine temporal shifts and how publications rely on pivotal or existing work.

\paragraph{Entity analysis}
We tracked the evolution of named entities in the \ac{esg}/\ac{dlt} content density filtered citation network, consolidating variations of similar entities (e.g., unifying all forms of \enquote{\acl{pow}} under \enquote{\acs{pow}}) using lemmatization and programmatic grouping to accurately capture changes in entity prevalence.

\section{Evaluation}

\subsection{Taxonomy labeling result}\label{subsec:LabelingResult}

Our \ac{ner} dataset organizes 39,427 named entities\footnote{An entity is counted from its beginning (\blockquote{B-}) to its end (\blockquote{I-}) using the IOB2 format \citep{Tjong1999RepresentingChunks}. See \autoref{tab:taskDLTexamples} for some examples.} from 80 manually labeled publications into a tree structure with 136 labels under 12 top-level categories (\autoref{fig:taxonomy}a and \autoref{tab:entityCounts}). This structure facilitated a targeted analysis in our study. \autoref{tab:taskDLTexamples} provides examples from the dataset.

\begin{table}[hbt]
\centering
\footnotesize %
\caption{Labeled named entities for each category in the dataset.}
\label{tab:entityCounts}
\setlength{\tabcolsep}{3pt} %
\renewcommand{\arraystretch}{1} %
\begin{tabular}
{
  l 
  S[table-format=5.0, table-number-alignment=center] %
}
\toprule

{\textbf{Entity Category}} & {\textbf{Number of Entities}} \\
\midrule
\rowcolor{LightGray} \texttt{Blockchain Name}                            & 4030 \\
\texttt{Consensus}                                    & 14530 \\
\rowcolor{LightGray} \texttt{Transaction Capabilities}                    & 5120 \\
\texttt{Native Currency Tokenization}               & 1382 \\
\rowcolor{LightGray} \texttt{Extensibility}                                & 1130 \\
\texttt{Security Privacy}                           & 5892 \\
\rowcolor{LightGray} \texttt{Codebase}                                    & 1251 \\
\texttt{Identity Management}                         & 1032 \\
\rowcolor{LightGray} \texttt{Charging \& Rewarding System}                   & 794 \\
\texttt{Identifiers}                                  & 1058 \\
\rowcolor{LightGray} \texttt{ESG}                                          & 2068 \\
\texttt{Miscellaneous}                               & 1140 \\

\bottomrule

\end{tabular}
\end{table}

\begin{table}[!htb]
\centering
\footnotesize %
\caption{Training examples showing text spans with their labels (\blockquote{Output}) and corresponding IOB2 format \citep{Tjong1999RepresentingChunks}, where \blockquote{B-} marks the beginning of an entity, \blockquote{I-} its continuation, and \blockquote{O} denotes non-entity tokens, used for fine-tuning \ac{bert}-based models for our \ac{ner} task.}
\label{tab:taskDLTexamples}
\begin{tabular}{p{0.95\textwidth} p{0.95\textwidth}}
\toprule

\textbf{Text:} \textit{In this paper, the PoW consensus algorithm used in blockchains are analyzed in terms of difficulty, hash count, and probability of successful mining.} \\ 
\textbf{Output:} \textit{In this paper, the \textcolor{blue}{$\langle$Consensus$\rangle$} consensus algorithm used in \textcolor{blue}{$\langle$Identifiers$\rangle$} is analyzed in terms of \textcolor{blue}{$\langle$Consensus$\rangle$}, \textcolor{blue}{$\langle$Transaction Capabilities$\rangle$}, and \textcolor{blue}{$\langle$Transaction Capabilities$\rangle$}.}\\
\textbf{IOB2:} \textit{O O O O \textcolor{blue}{$\langle$B-Consensus$\rangle$} O O O O \textcolor{blue}{$\langle$B-Identifiers$\rangle$} O O O O O \textcolor{blue}{$\langle$B-Consensus$\rangle$} O \textcolor{blue}{$\langle$B-Transaction\_Capabilities$\rangle$} O \textcolor{blue}{$\langle$B-Transaction\_Capabilities$\rangle$}}\\

\midrule

\textbf{Text:} \textit{Given the fundamental challenges in uniting Bitcoin mining with renewable energy, along with the fact that energy use is not the only way in which Bitcoin impacts the environment} \\ 
\textbf{Output:} \textit{Given the fundamental challenges in uniting \textcolor{blue}{$\langle$Consensus$\rangle$} with \textcolor{blue}{$\langle$ESG$\rangle$}, along with the fact that \textcolor{blue}{$\langle$ESG$\rangle$} is not the only way in which \textcolor{blue}{$\langle$Blockchain Name$\rangle$} \textcolor{blue}{$\langle$ESG$\rangle$}} \\
\textbf{IOB2:} \textit{O O O O O O \textcolor{blue}{$\langle$B-Consensus$\rangle$} O \textcolor{blue}{$\langle$B-ESG I-ESG$\rangle$} O O O O O \textcolor{blue}{$\langle$B-ESG$\rangle$} O O O O O O O \textcolor{blue}{$\langle$B-Blockchain\_Name$\rangle$} \textcolor{blue}{$\langle$B-ESG$\rangle$}}\\

\midrule

\textbf{Text:} \textit{the Hedera Treasury will \enquote{proxy-stake} over two-thirds of the total number of hbars to nodes hosted by Council Members.} \\ 
\textbf{Output:} \textit{the \textcolor{blue}{$\langle$Extensibility$\rangle$} will \enquote{\textcolor{blue}{$\langle$Consensus$\rangle$}} over \textcolor{blue}{$\langle$Consensus$\rangle$} of the total number of \textcolor{blue}{$\langle$Native Currency Tokenization$\rangle$} to nodes hosted by \textcolor{blue}{$\langle$Extensibility$\rangle$}.}\\
\textbf{IOB2:} \textit{O \textcolor{blue}{$\langle$B-Extensibility I-Extensibility$\rangle$} O \textcolor{blue}{$\langle$B-Consensus I-Consensus$\rangle$} O \textcolor{blue}{$\langle$B-Consensus$\rangle$} O O O O O \textcolor{blue}{$\langle$B-Native\_Currency\_Tokenization$\rangle$} O O O O \textcolor{blue}{$\langle$B-Extensibility I-Extensibility$\rangle$}}\\

\midrule

\textbf{Text:} \textit{Using a variant of Shor's algorithm [162], a quantum computer can easily forge an elliptic curve signature that underpins the security of each transaction in blockchain and so breaking of ECC  will affect blockchain in terms of broken keys, hence, digital signatures.} \\ 
\textbf{Output:} \textit{Using a variant of \textcolor{blue}{$\langle$Security Privacy$\rangle$} [162], a \textcolor{blue}{$\langle$Miscellaneous$\rangle$} can easily forge an \textcolor{blue}{$\langle$Security Privacy$\rangle$} that underpins the security of each transaction in \textcolor{blue}{$\langle$Consensus$\rangle$} and so breaking of \textcolor{blue}{$\langle$Security Privacy$\rangle$}  will affect \textcolor{blue}{$\langle$Consensus$\rangle$} in terms of \textcolor{blue}{$\langle$Security Privacy$\rangle$}, hence, \textcolor{blue}{$\langle$Security Privacy$\rangle$}.}\\
\textbf{IOB2:} \textit{O O O O \textcolor{blue}{$\langle$B-Security\_Privacy I-Security\_Privacy$\rangle$} O O \textcolor{blue}{$\langle$B-Miscellaneous I-Miscellaneous$\rangle$} O O O O \textcolor{blue}{$\langle$B-Security\_Privacy I-Security\_Privacy$\rangle$} O O O O O O O O \textcolor{blue}{$\langle$B-Consensus$\rangle$} O O O O \textcolor{blue}{$\langle$B-Security\_Privacy$\rangle$} O O \textcolor{blue}{$\langle$B-Consensus$\rangle$} O O O \textcolor{blue}{$\langle$B-Security\_Privacy$\rangle$} O \textcolor{blue}{$\langle$B-Security\_Privacy$\rangle$}}\\

\bottomrule
\end{tabular}
\end{table}

\subsection{\ac{ner} task result}\label{subsec:NERtaskResult}

\begin{table*}[htb!]
\centering
\footnotesize %
\caption{Detailed performance results (using seqeval \citep{Massias1999DesignRequirement,Nakayama2018Seqeval:Evaluation}) after 5-fold cross-validation fine-tuning for BERT, Albert, DistilBERT, and SciBERT with the ESG/DLT NER dataset, with relaxed and strict (exact) matched entities-level metrics.}
\label{table:performanceAllModelsComplete}

\renewcommand{\arraystretch}{1.2} %
\begin{tabularx}{\textwidth}{>{\hsize=.6\hsize\centering\arraybackslash}X 
                             >{\hsize=0.4\hsize\centering\arraybackslash}X 
                             *{8}{S[table-format=1.4]}}
\toprule
\rowcolor{white}
\multicolumn{2}{c}{\textbf{Model \& Fold}} & \multicolumn{4}{c}{\textbf{Relaxed Entity-Level Metrics}} & \multicolumn{4}{c}{\textbf{Strict Entity-Level Metrics}} \\
\cmidrule(lr){1-2} \cmidrule(lr){3-6} \cmidrule(lr){7-10}
\rowcolor{white}
\textbf{Model} & \textbf{Fold} & {\textbf{Precision}} & {\textbf{Recall}} & {\textbf{F1}} & {\textbf{Accuracy}} & {\textbf{Precision}} & {\textbf{Recall}} & {\textbf{F1}} & {\textbf{Accuracy}} \\
\midrule
\multirow{6}{*}{BERT} & 1 & 0.23 & 0.41 & 0.30 & 0.94 & 0.27 & 0.40 & 0.33 & 0.94 \\
 & 2 & 0.63 & 0.63 & 0.63 & 0.97 & 0.74 & 0.62 & 0.67 & 0.97 \\
& 3 & 0.45 & 0.48 & 0.47 & 0.96 & 0.59 & 0.46 & 0.52 & 0.96 \\
 & 4 & 0.54 & 0.50 & 0.52 & 0.96 & 0.60 & 0.49 & 0.54 & 0.96 \\
 & 5 & 0.64 & 0.79 & 0.70 & 0.97 & 0.70 & 0.78 & 0.74 & 0.97 \\
 & \textbf{Mean} &  0.50  &  0.56  &  0.52  &  0.96  &  0.58  &  0.55  &  0.56  &  0.96  \\
\midrule
\multirow{6}{*}{\centering Albert} & 1 & 0.30 & 0.31 & 0.30 & 0.94 & 0.34 & 0.31 & 0.32 & 0.95 \\
& 2 & 0.49 & 0.50 & 0.50 & 0.96 & 0.57 & 0.49 & 0.53 & 0.96 \\
 & 3 & 0.40 & 0.43 & 0.41 & 0.96 & 0.50 & 0.42 & 0.46 & 0.96 \\
& 4 & 0.76 & 0.80 & 0.78 & 0.98 & 0.81 & 0.79 & 0.80 & 0.98 \\
 & 5 & 0.80 & 0.83 & 0.81 & 0.98 & 0.85 & 0.83 & 0.84 & 0.98 \\
 & \textbf{Mean} &  0.55 &  0.57 &  0.56 &  0.97 &  0.61 &  0.57 &  0.59 & 0.97 \\
\midrule
\multirow{6}{*}{DistilBERT} & 1 & 0.26 & 0.38 & 0.31 & 0.93 & 0.31 & 0.37 & 0.34 & 0.93 \\
 & 2 & 0.57 & 0.62 & 0.59 & 0.97 & 0.65 & 0.62 & 0.63 & 0.97 \\
& 3 & 0.52 & 0.66 & 0.58 & 0.96 & 0.65 & 0.65 & 0.65 & 0.96 \\
 & 4 & 0.83 & 0.89 & 0.86 & 0.99 & 0.89 & 0.89 & 0.89 & 0.99 \\
 & 5 & 0.62 & 0.80 & 0.70 & 0.97 & 0.72 & 0.80 & 0.76 & 0.97 \\
& \textbf{Mean} &  0.56 &  0.67 &  0.61 &  0.97 &  0.64 &  0.67 &  0.65 &  0.97 \\
\midrule
\multirow{6}{*}{SciBERT} & 1 & 0.29 & 0.42 & 0.34 & 0.95 & 0.31 & 0.41 & 0.36 & 0.95 \\
& 2 & 0.63 & 0.66 & 0.65 & 0.98 & 0.71 & 0.66 & 0.68 & 0.98 \\
 & 3 & 0.50 & 0.63 & 0.55 & 0.96 & 0.62 & 0.62 & 0.62 & 0.96 \\
& 4 & 0.76 & 0.86 & 0.80 & 0.98 & 0.84 & 0.86 & 0.85 & 0.98 \\
 & 5 & 0.82 & 0.90 & 0.86 & 0.99 & 0.88 & 0.90 & 0.89 & 0.99 \\
 & \textbf{Mean} & 0.60 & 0.69 & 0.64 & 0.97 & 0.67 & 0.69 & 0.68 & 0.97 \\
\bottomrule
\end{tabularx}
\end{table*}

We fine-tuned four pre-trained transformer models: the \ac{bert} base model (cased)\footnote{https://huggingface.co/bert-base-cased}, which uses a bidirectional encoder architecture \citep{Devlin2019BERT:Understanding} (see \autoref{subsec:ner-literature-mining} for more details); ALBERT base model v2\footnote{https://huggingface.co/albert-base-v2}, a lighter version of \ac{bert} with a cross-layer parameter sharing technique \citep{Lan2019ALBERT:Representations}; DistilBERT base model (cased)\footnote{https://huggingface.co/distilbert-base-cased}, a distilled version of \ac{bert} \citep{Sanh2019DistilBERTLighter}; and SciBERT base model (cased)\footnote{https://huggingface.co/allenai/scibert\_scivocab\_cased}, a \ac{bert}-based model pre-tained on a corpus of scientific publications \citep{Beltagy2019SciBERT:Text}. To ensure robust evaluation and prevent data leakage, we employed 5-fold cross-validation based on publication titles, ensuring that data from any single publication appeared exclusively in the training or test set for each fold. The training process consisted of 200 total epochs, distributed as 40 epochs per fold, with a learning rate of \(5 \times 10^{-5}\), a training batch size of 32, and a validation batch size of 64. The maximum sequence length was 512 tokens.

The performance evaluation (\autoref{table:performanceAllModelsComplete}) revealed that SciBERT \citep{Beltagy2019SciBERT:Text} achieved the highest F1 score for our domain-specific \ac{ner} task in \ac{dlt}. This aligns with expectations, given SciBERT's pre-training on a large multi-domain corpus of scientific literature, which improves its performance in downstream scientific \ac{nlp} tasks \citep{Beltagy2019SciBERT:Text} like ours. Despite SciBERT's results, we ultimately chose DistilBERT, the second-best-performing model (\autoref{table:performanceAllModelsComplete}), for processing our large dataset of 24,539 publications. DistilBERT offers a reasonable balance of effectiveness and efficiency, operating 60\% faster than BERT, and likewise SciBERT, during inference while retaining 97\% of BERT's performance \citep{Sanh2019DistilBERTLighter}. This speed and F1 score combination made DistilBERT the most suitable choice for our large-scale analysis of \ac{dlt} literature.

\section{Discussion}
\subsection{Transferability and adaptability of the methodology}
\subsubsection{Methodology framework transferability}

Our methodology shows versatility in analyzing emerging technologies and their interdisciplinary implications. While we applied it to \ac{dlt}\footnote{\autoref{glossary} explains \ac{dlt} and some its key terms in more detail} and \ac{esg}, the framework's structure makes it adaptable to various technological domains and their intersections, such as \ac{ai} in healthcare or renewable energy applications in urban development, to name a few.

\paragraph{Methodological framework}
The methodology comprises five phases (\autoref{fig:methodology}). Initially, we established a comprehensive corpus through data collection by selecting seed papers that represent seminal publications at domain intersections of interest via citation analysis and expert consultation, then systematically expanding this by traversal up to two levels of the seed paper's references (\autoref{subsec:DataCollection}). The second phase focused on data labeling by first developing taxonomies that capture the domain intersections terminology, which serves as a guideline for labeling a \ac{ner} dataset (\autoref{subsubsec:labeling} and \autoref{subsubsec:LabelsProcessing}). This labeled dataset then supported the third phase in which it is used to fine-tune \ac{bert}-based models for a \ac{ner} task (\autoref{subsec:ner-literature-mining}), benchmarking and selecting the best-performing model (\autoref{subsec:NERtaskResult}). The fourth phase applied the selected fine-tuned model to filter the corpus based on the named entities identified (\autoref{subsubsec:DatasetFiltering}). Finally, we conducted a manual literature review and temporal graph analysis to derive insights that track the evolution of concepts, technologies, and their interconnections.

\paragraph{Framework application example}
To demonstrate the framework's transferability, consider its application to \ac{ai} in healthcare research. The process would begin by identifying influential papers at the \ac{ai}-healthcare intersection, followed by traversal of the references from these papers to build a comprehensive corpus. A specialized taxonomy would integrate medical terminology with \ac{ai} concepts, enabling dataset labeling for model fine-tuning. This fine-tuned model with the labeled dataset would then facilitate the identification of specific medical \ac{ai} technologies and their clinical applications. Finally, a manual literature review using temporal graph analysis, named entities, and citation network analysis would reveal the evolution of medical \ac{ai} research and its healthcare impact.

\subsubsection{Methodological advantages over other approaches}

Our methodology (\autoref{sec:methodology}) of using \ac{ner} for field-specific literature filtering is similar to recent \ac{llm} developments, such as Google DeepMind Gemini's demonstration of a systematic literature review. In Gemini's approach, terms like \enquote{Chip} and \enquote{CRISPR-Cas9} are searched in publication titles and abstracts to filter results\footnote{See \url{https://youtu.be/sPiOP_CB54A?feature=shared&t=64} for the prompt used in the demonstration}. This parallel highlights the significance of our work. However, Gemini faces limitations like potential hallucinations that could undermine it for \ac{ner} tasks and filtering of publications. Although supervised learning approaches outperform current \acp{llm} for \ac{ner} tasks \citep{Li2023AreTasks,Hu2024ImprovingEngineering}, Gemini shows the potential of \acp{llm} in few-shot learning for \ac{ner} tasks.

In contrast to current \ac{llm} approaches, our methodology (\autoref{sec:methodology}) uses a domain-specific labeled \ac{ner} dataset and a fine-tuned pre-trained language model to analyze full-text publications. This supervised learning approach provides a more structured and verifiable framework for literature analysis, making it particularly valuable for systematically examining emerging technical fields and their intersections with other domains. The methodology's emphasis on supervised learning and domain-specific training ensures greater accuracy and reliability (see \autoref{subsec:ner-literature-mining} for more details) in literature review tasks while maintaining adaptability to various fields of study.

\subsection{Findings}

\subsubsection{Keywords network}\label{subsec:TopicsNetwork}

\begin{figure}[htb]
   \centering
     \centering
          \includegraphics[width=0.45\textwidth]{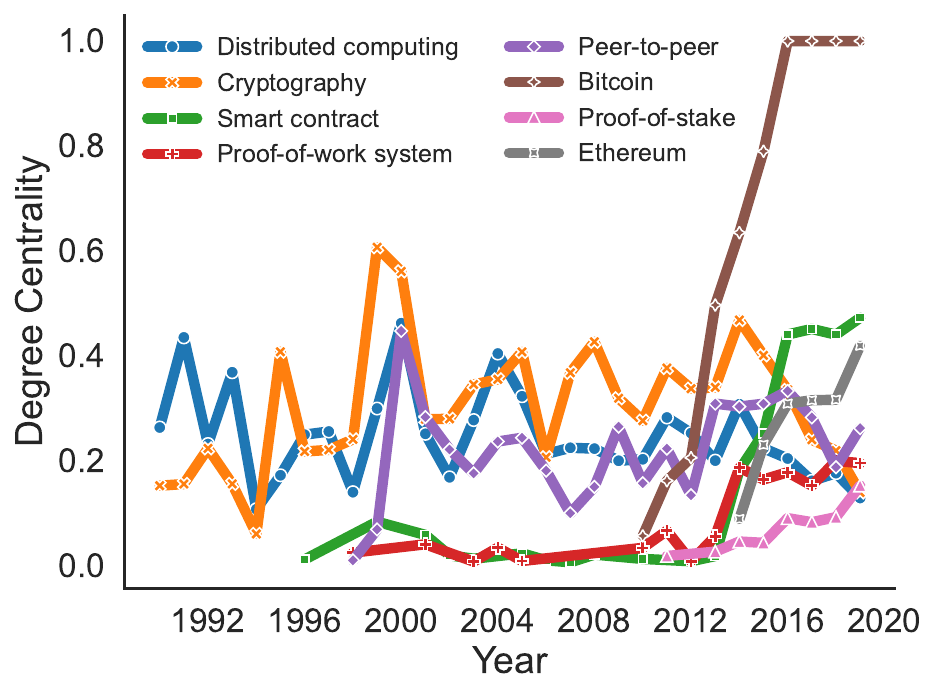}
     \captionsetup{size=footnotesize}
     \caption{Evolution of research topics in the keywords graph from the full corpus of 63,096 publications.}
     \label{fig:topicsGrowth}
\end{figure}

The keywords graph comprises 25,048 topics as nodes with 3,042,397 edges and 4,847,870 temporal edges connecting the topics across time. \autoref{fig:topicsGrowth} visually and quantitatively shows the evolution of the key areas like Cryptography, Peer-to-Peer, \ac{pow}, \ac{pos}, and smart contracts that laid the foundation for advancements in \ac{dlt} at different points in time. The research in Cryptography provided the foundations for using public and private key pairs for identity management \citep{Chaum1981UntraceablePseudonyms}, elliptic curves \citep{Miller1986UseCryptography}, hashing, and Merkle trees \citep{Merkle1980ProtocolsCryptosystems}. Similarly, Peer-to-Peer research focused on network communication \citep{Cohen2003IncentivesBitTorrent,Dingledine2004Tor:Router}, timestamping \citep{Haber1991HowDocument,Bayer1993ImprovingTime-Stamping,Massias1999DesignRequirement}, and the need for solving the Byzantine Generals Problem \citep{Lamport1982TheProblem}, vital for decentralized network functionality. The research related to \acl{pow} delineated the first consensus system used in Blockchain building in the works from the early 1990s of \citet{Dwork1993PricingMail} and the early 2000s of \citet{Back2002HashcashCounter-Measure,Finney2004ReusableRPOW}. 

The emergence of Bitcoin \citep{Nakamoto2008Bitcoin:System} marked a significant convergence of these technologies. Its degree centrality as a research topic rapidly soared from 0.056 in the late 2010s to 1 in the 2020s, a growth of $\sim$1,692.42\%. This surge in interest, possibly fueled by the socio-economic and political climate following the 2008 financial crisis \citep{DavidYaffe-Bellany2023HasTimes,GautamMukunda2018TheLater,Zamani2017TheGreece,Zewde2022ImpactCohorts,Schepisi2021TheReview} and Bitcoin's technological innovation, contrasted with digital money projects \citep{Chaum1996DavidCEO,Dai1998B-money,Szabo2005BitGold} in the 1990s and early 2000s that did not become as popular as Bitcoin in the same amount of time.

Parallel to Bitcoin's rise, or inspired by it \citep{Cornish2018EthereumsBlockchain}, Ethereum emerged around 2014-2015, introducing two key technological innovations: the Solidity programming language and the \ac{evm} for smart contract development and execution, respectively \citep{Buterin2014Ethereum:Platform.}. Ethereum's significance in the field grew substantially, with its degree centrality increasing from 0.088 in 2014 to 0.419 in 2020s (a $\sim$376.64\% growth). While smart contracts were conceptualized in the mid-1990s \citep{Szabo1997FormalizingNetworks}, their adoption remained limited until Ethereum's implementation \citep{Buterin2014Ethereum:Platform.}, which provided a comprehensive development and execution environment through Solidity and the \ac{evm}. This technological foundation catalyzed extraordinary growth in smart contract research interest, evidenced by a degree centrality increase from 0.011 in the mid-1990s to 0.472 in the 2020s (a $\sim$4,299.04\% growth). This research interest in smart contracts is reflected in efforts to extend its applicability, including legal aspects through Smart Legal Contracts \citep{Roche2021ErgoContracts} and their generation using \ac{nlp} \citep{Chen2023ConversionNLP}.

Amid these developments, \acl{pos} was introduced as an energy-efficient alternative to \acl{pow} by \citet{King2012PPCoin:Proof-of-Stake} in 2012, which grew from 0.017-degree centrality since its introduction to 0.152 in the 2020s, a $\sim$768.68\% growth, reflecting the growing interest in more sustainable consensus mechanisms.

\begin{figure*}[htb]
  \centering
    \input{figures/networkClusterLevels}
    \hfill
    \input{figures/networkESG}
    \captionsetup{size=footnotesize}
    \caption{ESG/DLT network derived from 505 papers filtered by their content density of the ESG/DLT thematic (\autoref{subsubsec:DatasetFiltering}). The network comprises 10,172 nodes (citations), 15,898 edges, and 20,111 temporal edges. Colors represent different layers (\autoref{fig:netClusterLevel}) or content density for \ac{esg} entities (\autoref{fig:netESGDensity}), node sizes indicate authority scores and only top nodes are displayed.}
    \label{fig:citationNetworks}
\end{figure*}

\subsubsection{ESG/DLT network} \label{subsec:CitationNetwork}

\begin{figure*}[tb]
  \centering
    \input{figures/filteredNetEvol}
    \hfill
    \input{figures/networkClusterYears}
    \captionsetup{size=footnotesize}
    \caption{ESG/DLT network's evolution in terms of publications (\autoref{fig:seedvsfiltered}) and time windows (\autoref{fig:netClusterYears}).}
    \label{fig:citationNetworkEvolution}
\end{figure*}

\begin{figure*}[bt]
\centering
    \input{figures/normalizedNamedEntitiesEvol}
    \hfill
    \input{figures/termsConsensus}
    \hfill
    \input{figures/termsESG}
    \captionsetup{size=footnotesize}
    \caption{Named entities evolution in the citation network.}
    \label{fig:tempGraphAnalysis}
\end{figure*}

The ESG/DLT content density citation graph, derived from 505 papers, comprises 10,172 nodes (each node representing a citation), 15,898 edges, and 20,111 temporal edges connecting the publications over time. 
This graph facilitates narrowing down the publications specific to the ESG/DLT intersection (see \autoref{fig:workflow} and \autoref{subsubsec:DatasetFiltering}). The ESG/DLT network (\autoref{fig:netClusterLevel}) reveals a distinctive pattern in research evolution, particularly evident in the second layer of references, where multidisciplinary research demonstrates increased prevalence (\autoref{fig:netClusterLevel} and \autoref{fig:netESGDensity}). This layer exhibits a notable concentration of \ac{esg}-focused investigations, especially regarding energy consumption and efficiency optimization of blockchain systems (\autoref{fig:netESGDensity}). The peripheral references display an even broader interdisciplinary scope, which extends beyond traditional computer science boundaries into domains such as finance and environmental studies. This is likely because, as the \ac{dlt} field develops, interdisciplinary research emerges. 

The network (\autoref{fig:seedvsfiltered} and \autoref{fig:netClusterYears}) shows a publication surge between 2008 and 2011 of $\sim$191.03\%, aligning with Bitcoin's release and its subsequent influence on diverse \ac{dlt} research areas, particularly in consensus mechanisms (\autoref{fig:labelsDensity}\footnote{The \texttt{Blockchain Name} entities identified before 2007, such as HashCash \citep{Back2002HashcashCounter-Measure}, reflect the standard industry practice of using \enquote{Blockchain} as a generic term for any consensus-based \ac{dlt} system, a convention we maintain in our taxonomic classification.}). Notably, Bitcoin's persistent and increasing significance in research attention (see also \autoref{subsec:TopicsNetwork}), despite the emergence of alternative solutions, demonstrates a \enquote{Lindy effect} \citep{Goldman1964LindysLaw, Mandelbrot1983TheNature}, where the protocol's longevity correlates with its continued academic interest. As Bitcoin gained more public attention after its appearance in 2008, often driven by price action, the research community evaluated this new technology, particularly Bitcoin's \ac{pow} \citep{Vukolic2016TheReplication,Biryukov2015Proof-of-workRelay}, for potential unaccounted environmental costs \citep{deVries2018BitcoinsProblem,Mora2018Bitcoin2C} or negative externalities \citep{Jones2022EconomicGold,Papp2023BitcoinDecisions}.

Despite Bitcoin's prominent media coverage regarding energy consumption, our analysis of the ESG/DLT citation network (\autoref{fig:citationNetworks} and \autoref{fig:labelsDensity}) reveals two significant patterns in research focus. First, ESG-related research has maintained a consistent proportional representation within the broader DLT literature rather than showing increased relative dominance (\autoref{fig:labelsDensity}), with its primary focus remaining on energy consumption and efficiency optimization of blockchain systems (\autoref{fig:netESGDensity}). Second, the research emphasis has evolved from an early focus on fundamental security and privacy considerations (\autoref{subsec:TopicsNetwork}) toward a growing focus on efficient \citep{King2012PPCoin:Proof-of-Stake} and secure consensus algorithms \citep{Douceur2002TheAttack}, and blockchain architectures \citep{Vukolic2016TheReplication,Bonneau2015SoK:Cryptocurrencies}. Curiously, we observe an increase in these studies investigating energy-efficient consensus mechanisms and scaling solutions during periods of rapid price appreciation, such as Bitcoin's bull runs \citep{2024CryptoBullRun,Canellis2023BitcoinBlockworks,Lang2024Cryptoverse:Reuters}.

Post-2012, the network saw a marked increase in publications, especially after 2014, reflecting the impact of seminal works like PPCoin and Ethereum's whitepapers \citep{King2012PPCoin:Proof-of-Stake, Buterin2014Ethereum:Platform.}. The growth in citations and publications from 2012 until the 2020s was $\sim$901.75\%. The increased interest in \ac{pos} as an energy-efficient alternative to \ac{pow} \citep{Perez2020} is exemplified by Vitalik Buterin's early exploration of \ac{pos} before Ethereum's launch using \ac{pow} in 2015, which is evident in his 2014 Slasher algorithm post \citep{Buterin2014Slasher:Algorithm} and later posts that discuss \ac{pos} benefits \citep{Buterin2014OnStake}, and the \enquote{nothing at stake} challenge \citep{Buterin2014ProofSubjectivity}.

Vitalik's early public discussions of \ac{pos} may have been one of the catalysts that encouraged further research into \ac{pos} (\autoref{fig:topicsGrowth} and \autoref{fig:termsConsensusEvol}) before Ethereum's transition from \ac{pow} to \ac{pos} in 2022 \citep{EthereumFoundation2022TheMerge}. This is consistent with how Ethereum popularized smart contracts despite the research around smart contracts existing since the mid-1990s \citep{Szabo1997FormalizingNetworks}, but remaining seemingly static as a topic of interest for the research community (\autoref{fig:topicsGrowth}) until after Ethereum introduced the Solidity programming language and \ac{evm} (\autoref{subsec:TopicsNetwork} and \autoref{fig:topicsGrowth}).

\ac{dlt}'s thematic shift, coupled with the increasing prominence of terms like \enquote{decentralization}, \enquote{blockchain}, and \enquote{sustainability}, underscores a multidisciplinary approach in the field (\autoref{fig:termsESGEvol}). The sustained interest in \ac{pow}, along with explorations into \ac{pos} and other consensus mechanisms, highlights the field's adaptability to environmental and scalability challenges (\autoref{fig:termsConsensusEvol}).

\subsection{Limitations} \label{subsec:limitations}

Our literature review faces limitations, including potential biases in seed paper selection and a time lag in capturing recent publications, which may affect the comprehensiveness of our analysis. For instance, the choice of XRP's 2018 whitepaper \citep{Chase2018AnalysisProtocol} over the more cited 2014 edition \citep{Schwartz2014TheAlgorithm} could underestimate its influence on the citation network. Similarly, recent works such as the 2018 Hedera whitepaper \citep{Baird2018Hedera:Council} are omitted due to unavailable citation data.

Building the citation network predominantly from pre-2020 seed papers introduces a bias toward older publications, potentially overlooking newer research yet to achieve recognition (\autoref{fig:seedvsfiltered}). While our methodology could theoretically filter citations to seed papers based on content density, our review focused solely on references within the seed papers, possibly limiting the thematic breadth.

Regarding our \ac{ner} dataset performance, there is an expected noise level \citep{Fetahu2023MultiCoNERRecognition,Alfina2017ModifiedDataset,Rucker2023CleanCoNLL:Dataset} that we tried to reduce by enforcing interlabeler consistency (see \autoref{subsubsec:LabelsProcessing}). Our dataset, with 80 full-text manually labeled papers, is smaller than many general-purpose \ac{ner} datasets \citep{Weischedel2017OntoNotesProcessing,Tjong2003IntroductionRecognition,Tedeschi2021WikiNEuRal:NER,Loukas2022FiNER:Tagging} but is comparable to those in specialized domains like materials science \citep{Cheung2024POLYIE:Literature,Dagdelen2024StructuredModels}.

Our analysis contains 24,539 openly accessible publications from an initial corpus of 63,083 references (\autoref{fig:workflow}). While this reliance on publicly available research introduces potential sampling limitations, several factors mitigate concerns about representativeness and comprehensiveness. First, empirical evidence demonstrates that open-access publications achieve broader academic engagement \citep{Piwowar2018TheArticles,Archambault2014ProportionAccess,McCabe2014IDENTIFYINGJOURNALS} and higher media attention \citep{Schultz2021AllMedia}, suggesting they effectively capture significant research developments and may influence more research trends. Second, although only approximately 18\% of journal articles are currently openly available \citep{Piwowar2018TheArticles}, high-quality research authors increasingly seem to favor hybrid open-access journals \citep{Gaule2011GettingHelp}. Third, the \ac{dlt} field's rapid evolution has elevated non-traditional literature, particularly whitepapers and industry publications, as crucial sources of technological developments. Seminal works such as the whitepapers for Bitcoin \citep{Nakamoto2008Bitcoin:System}, Ethereum \citep{Buterin2014Ethereum:Platform.}, and PPCoin \citep{King2012PPCoin:Proof-of-Stake} exemplify this trend. Our methodology addresses these considerations through two complementary approaches: (1) incorporating non-traditional academic literature to capture contemporary technological developments, and (2) leveraging metadata from paywalled publications in our broader analysis of the \ac{dlt} field evolution (\autoref{subsec:TopicsNetwork}). This two-tier analytical approach (\autoref{sec:methodology}) ensures comprehensive coverage of \ac{dlt} research developments while maintaining methodological rigor.

\subsection{Future work}
Future work, as outlined in \autoref{subsec:limitations}, should focus on integrating metadata from different whitepaper versions, like XRP's 2014 and 2018 editions \citep{Schwartz2014TheAlgorithm,Chase2018AnalysisProtocol}, and sourcing metadata from alternative databases for publications with missing information, such as Hedera's whitepaper \citep{Baird2018Hedera:Council}. Additional research should also include regular updates to the taxonomy (refer to \autoref{tab:labelsDLT}), expanding training data by annotating more seed papers and exploring various language model architectures.

Additionally, beyond the systematic literature review presented in this paper, our \ac{ner} dataset could also support other cross-domain literature review research, such as \ac{dlt} applications across software development, cybersecurity, and governance frameworks. Also, it could enable relationship mapping between \ac{dlt} concepts, supporting the construction of knowledge graphs. Finally, the dataset could contribute to developing automated systematic literature review systems, particularly valuable given the rapid evolution of the \ac{dlt} field. These potential applications emphasize our \ac{ner} dataset's versatility and value beyond this paper.

\section{Conclusion} \label{sec:conclusions}

This paper addresses the scarcity of high-quality labeled \ac{nlp} data for blockchain research by developing an \ac{ner} dataset for the \ac{esg}/\ac{dlt} domains intersection from public sources. We conducted a systematic literature review analysis to demonstrate its utility and provide future research possibilities. Our analysis reveals the critical intersections of topics like Cryptography, Peer-to-Peer, \acl{pow}, \acl{pos}, and smart contracts in the evolution of \ac{dlt}. We observed Bitcoin's persistent dominance in research (a \enquote{Lindy effect}). Ethereum's significant impact on the adoption of smart contracts and \ac{pos} as an energy-efficient alternative to \ac{pow}. These insights, and the growing focus on energy-efficient consensus mechanisms, highlight the field's rapid adaptation to technological and environmental challenges.

We believe our work represents a step towards leveraging and facilitating the use of \ac{nlp} to research rapidly evolving fields like \ac{dlt}, enabling researchers, policymakers, and industry stakeholders to stay informed and make data-driven decisions.

\section*{Acknowledgment}

We thank Ali Irzam Kathia for labeling a subset of the \ac{ner} dataset. We also thank Editha Nemsic for her contribution to the early stage of the study.

\section*{Author contributions}
Walter Hernandez Cruz: Conceptualization, Data curation, Formal analysis, Investigation, Methodology, Project administration, Software, Validation, Visualization, Writing—original draft, Writing—review \& editing. Kamil Tylinski: Conceptualization, Data curation, Methodology, Visualization, Writing—original draft, Writing—review \& editing. Alastair Moore: Conceptualization, Data curation, Methodology, Visualization, Writing—original draft, Writing—review \& editing. Niall Roche: Data curation, Methodology, Software, Writing—review \& editing. Nikhil Vadgama: Conceptualization, Data curation, Methodology, Resources, Writing—review \& editing. Horst Treiblmaier: Conceptualization, Data curation, Methodology, Writing—review \& editing. Jiangbo Shangguan: Data curation, Software. Paolo Tasca: Conceptualization, Validation, Supervision, Resources, Writing—review \& editing. Jiahua Xu: Conceptualization, Validation, Supervision, Resources, Writing—review \& editing.

\section*{Competing interests}
The authors have no competing interests.

\section*{Funding information}
This research received no external funding.

\section*{Data availability}
The \ac{ner} dataset generated during this study is available at \url{https://huggingface.co/datasets/ExponentialScience/ESG-DLT-NER}

\clearpage
\glsaddall
\printglossary[nonumberlist] \label{glossary}

\clearpage

\printbibliography

\end{document}